\documentclass[11pt,review]{elsarticle}
\biboptions{numbers,sort&compress} 
\journal{arXiv}
\usepackage{hyperref}

\usepackage[utf8]{inputenc}
\usepackage{graphicx}
\usepackage{booktabs} 
\usepackage{xcolor}
\usepackage{amsmath, amssymb, amsthm} 
\allowdisplaybreaks

\makeatletter
\let\@afterindenttrue\@afterindentfalse
\makeatother

\begin{document}

\begin{frontmatter}

\title{A note on the logical inconsistency of the Hotelling Rule: A Revisit from the System's Analysis Perspective}

\author[addr-iiasa-em]{Nikolay Khabarov\corref{tag-correspondingauthor}}
\cortext[tag-correspondingauthor]{Corresponding author}
\ead{khabarov@iiasa.ac.at} 

\author[addr-iiasa-ibf,addr-msu]{Alexey Smirnov}
\author[addr-iiasa-em,addr-oxford]{Michael Obersteiner}

\address[addr-iiasa-em]{Exploratory Modeling of Human-Natural Systems Research Group, 
Advancing Systems Analysis Program,
International Institute for Applied Systems Analysis (IIASA), 
Schlossplatz 1, Laxenburg, A-2361, Austria}

\address[addr-iiasa-ibf]{Integrated Biosphere Futures Research Group,  Biodiversity and Natural Resources Program,
International Institute for Applied Systems Analysis (IIASA), 
Schlossplatz 1, Laxenburg, A-2361, Austria}

\address[addr-msu]{Faculty of Computational Mathematics and Cybernetics,
Lomonosov Moscow State University, Moscow, 119991, Russia}

\address[addr-oxford]{Environmental Change Institute, Oxford University Centre for the Environment, South Parks Road, Oxford, OX1 3QY, UK}

\begin{abstract}
The "Hotelling rule" (HR) called to be "the fundamental principle of the economics of exhaustible resources"~\cite{refSolow1974} has a logical deficiency which was never paid a sufficient attention to. This deficiency should be taken into account before attempting to explain discrepancies between the price prediction provided by the HR and historically observed prices. Our analysis is focused on the HR in its original form, we do not touch upon other aspects such as varying extraction costs and other ways of upgrading the original model underlying the Hotelling rule. We conclude that HR can not be derived from the simple models as it was claimed, therefore it should be understood as an assumption on its own, and not as a rule derived from other more basic assumptions.

	\phantom{space-for-keywords} 

\end{abstract}


\end{frontmatter}

\section{Introduction}

In his seminal paper~\cite{refHotelling1931} devoted to economics (and pricing) of exhaustible resources Harold Hotelling hypothesized that in a perfect competition case "it is not unreasonable to expect that the price $p$ will be a function of the time of the form $p = p_o \exp(\gamma t)$". While there are later studies providing plausible explanations e.g.~\cite{refNgoXXX}, Hotelling did not provide his own explanation in that seminal paper~\cite{refHotelling1931}. Solow~\cite{refSolow1974} provided a justification for selecting that specific price function, which we discuss below in the Section~A.

In that same paper~\cite{refHotelling1931} in Chapter 3 Hotelling provided a justification for the same exponential price path also for the case of the "maximum social value". We discuss this case in the Section~B of this manuscript.

\section{Section A}

The logic in a perfect competition case is that a company---a price taker--- wants to find an optimal resource extraction schedule to maximize its discounted profit from selling the resource over a time period when the price path is given i.e. the function $p(t)$ is known and fixed. It turns out that the solution of that problem exists only in the specific case if $p(t) = p_o \exp(\gamma t)$, where $\gamma$ is the market interest rate. As Solow comments~\cite{refSolow1974} on page 3, if the price would grow slower -- the company will aim at extracting all its quantity of resource immediately, if it would grow faster -- the company would aim at waiting until the very end of the modeling period and then extract at that moment everything. Since the extraction of the full quantity of the resource in a moment of time (instantaneously) is assumed to be infeasible, the specific case of  $p(t) = p_o \exp(\gamma t)$ is declared to be a solution. Solow in~\cite{refSolow1974} emphasizes that in that case the company is indifferent to extracting or not extracting any particular quantity of resource in any time moment within the planning time interval. 

There are three remarks to consider just before we conclude with the summary for this section:

a) the initial posed problem on finding an optimal extraction schedule was not solved by this consideration as there is no unique solution found, moreover, any resource extraction schedule would be a solution under the suggested price path, which in turn makes the initially posed problem meaningless;

b) to remove the problem highlighted by Solow~\cite{refSolow1974} and have a well-defined solution for any ad hoc fixed price path $p(t)$, one could alternatively assume that the quantity of the resource available to a company is extractable within one modeled time moment (decade, year, month, or week), which only requires to legitimately get rid of the mathematical abstraction on infinitely small length of the time subintervals where company makes its extraction plans; alternatively, a limit on the maximum feasible rate of extraction would achieve the same goal;

c) in the posed problem, the resource price $p(t)$ was supposed to be given, and if it was different from $p(t) = p_o \exp(\gamma t)$ then one has to live with that and accept $p(t)$ still to be valid and the problem to not have a solution. The situation when an economic problem does not to have a solution is not something extraordinary: even a very basic problem might not have a solution. Here is an example of that mere fact, disconnected from the resource extraction problem: there are two suppliers and each supplier has 2 apples and wants to sell them all at \$1 per unit price. A customer has \$2 and wants to buy two apples. How many apples would each supplier sell? Nobody can provide a definite solution: there are many "ifs" attached to it -- would the first supplier talk to the customer before the second has a chance? Does the customer want to buy two apples at once or, one apple in the morning and another one in the afternoon? There are other questions that one could ask and which are not part of the initial problem setting, whereas they could help to define a solution. In this situation, it would be illegitimate to instead assume that the price of the first supplier is actually \$4 per apple, in order to come up with a solution of a customer buying apples from the second supplier.

In summary, the assumption of the very specific case --- if $p(t) = p_o \exp(\gamma t)$ --- is providing an answer to a wrong question. While the original question was "what is the optimal extraction plan", the answer provided is "the price is X if the company should not care about the optimal extraction plan". We find this logic to be deficient as the initial driving question is completely disregarded in the answer.  

Of course, one could attempt to fix that logical deficiency: knowing the "answer" one could ask a "compatible" question: "what should be the market price if a company does not care whether to extract its resource now or in the future".
When seen from the systems analysis perspective, this question is posed at both macro-level (market price) and micro-level (no incentive for a company to extract resource). Because of inclusion of both macro and micro levels, this question now presents another logical deficiency: at the macro level the economy needs the resource to work and hence assumes incentives for companies at micro level to provide a supply at each time moment, whereas these incentives are denied by requiring a company to not care about supplying the resource at each time moment.
In other words, the question is posed to determine the market price, whereas the only specification of the market given is that nobody from the sellers has any incentives to sell. This specification of the "market" is by far not complete because (while the extraction cost of suppliers is known), the buyers on that market are not described in any way. 

So, in the perfect competition case the logical deficiency existing in the suggested solution $p(t) = p_o \exp(\gamma t)$ cannot be fixed by posing another problem, which is compatible with this suggested solution.

\section{Section B}

In the "maximum social value" case presented by Hotelling in the Chapter 3 of his seminal paper~\cite{refHotelling1931} he is deducing the same exponential form of the price path. 
Here the logic is different from the case of the perfect competition. 
First, the price for the incremental unit of resource is described by the demand curve $p=\tilde p(q)$, which is a diminishing function of the quantity $q$. 
Second, the social planner is supposed to maximize the total time-discounted utility, which at each time moment is represented as the consumer and producer surplus (assuming zero production/extraction cost for simplicity). 
Then the conclusion is made on the optimal price being $p(t) = p_o \exp(\gamma t)$.

We argue that in this "maximum social value" case the logic is broken, because, from the systems perspective,  economic social planner (by making decisions on resource extraction schedule) is shaping economic development.
Naturally, different achievabe levels of economic development imply different levels of demand for the resource, but the economic development dynamics is not included in the demand curve $\tilde p(q)$.

In an attempt to fix that logical deficiency, one has to include into the problem formulation a representation of economy that the social planner should take into account (and aparently regulate) when deciding on resource extraction schedule  aimed at maximization of social welfare. 
This effort has already been undertaken and the respective problem including economic dynamics was formulated by Dasgupta and Heal (see (1.6) in~\cite{refDasgupta1974}).
The application of the optimal control theory and the Pontryagin Maximum Principle to this problem (p.234 (10.59) in~\cite{xrefHritonenko}, p.199 (60) in~\cite{xrefWeitzman}) leads to consideration of co-state variables, where the one corresponding to the resource stock --- the shadow resource price --- is indeed described as $p(t) = p_o \exp(\gamma t)$.

While the shadow price has been declared as the optimal resource price, there is no justification for doing so.
The economic meaning of the shadow price is that it is the increment of the optimal total welfare corresponding to an increase in the availability of the constrained resource at a fixed time moment~\cite{xrefSydsater,xrefIntrilligator,xrefAseev}.
Naturally, the social planner would be willing to pay any amount below that price to increase available resource if that would be possible, because by doing so (and using additional amount of resource), the social welfare could be increased.
However, since the Dasgupta (and Hoteling) problem is formulated with the strict resource constraint, such possibility will never be seen by the social planner within the problem's context.
So, the social planner will not pay that price under any circumstances, and nobody else would, as the control is supposed to be all with the social planner.
As the shadow price is never paid and even cannot be paid within the problem's context, this value has no economic meaning and only serves as a technical means for obtaining the optimal resource extraction schedule.
"No economic meaning" means here also that this value cannot be seen as the resource market price.

In short, within the the social planner paradigm, for the social planner, at the whole economy scale, the cost of resource extraction (constant in HR) and the created total economic welfare (utility) are sufficient to define the optimal extraction schedule.
The path of the resource price --- the price actually paid from one economic sector to another --- has nothing to do with it.
The resource price is used internally by economy to trade between its sectors, so that price is irrelevant to the simple social planner paradigm (that does not even include any representation of the inter-sectoral trade).
Logically, one cannot derive the value of a parameter irrelevant to a paradigm from that paradigm.
Therefore HR cannot deliver economically meaningful resource price also in the "social planner" case --- even after attempting to fix the initial logic deficiency with the disconnect between the importance of the resource for economy and the absence of the economy representation in the initial optimization problem.

\section{Conclusion}

Logical analysis of HR carried out in sections A and B for the two problem settings -- perfect competition and social planner, from where the HR is originally derived -- has lead to a discovery of severe logical deficiencies.
Since one could hardly build "the fundamental principle of the economics" based on such a deficient logic as the HR is build upon, we suggest to discard HR as a "rule" completely.
In case of the perfect competition, HR should be understood as an assumption on its own, and not as a rule derived from other assumptions. 
In case of maximization of social value, HR should not be deemed as carrying any economic meaning.

\section*{Acknowledgments} 

The authors acknowledge early discussions (mostly centered around the DICE model~\cite{refDICEWebPage}) with their IIASA colleagues Elena Rovenskaya, Artem Baklanov, Fabian Wagner, Thomas Gasser, Petr Havlik, Armon Rezai, Michael Kuhn, Stefan Wrzaczek, Michael Freiberger, Johannes Bednar, and others.
These discussions indirectly spurred the interest in a deeper exploration of the Hotelling rule that ultimately resulted in the presented analysis.

\section*{Funding} 

Austrian Science Fund (FWF): P31796-N29/``Medium Complexity Earth System Risk Management'' (ERM).

\section*{Author contributions} 

NK has conceptualized the problem; NK and AS have carried out the investigation; MO has contributed to funding acquisition; NK, AS, and MO have discussed the results in the process of investigation; NK has drafted the paper; all co-authors have contributed to writing the manuscript.

\section*{Competing interests} 

Authors declare no competing interests. 

\section*{Data and materials availability} 

Not applicable for this theoretical study. 

\section*{Supplementary Materials}

None.


\end{document}